\documentclass[aps,pre,bibnotes,reprint,twoside,showpacs,preprintnumbers,amsmath,amssymb,citeautoscript]{revtex4-1}

\usepackage{graphicx}

\usepackage{bm}

\usepackage[utf8]{inputenc}

\usepackage{hyperref}

\newcommand{\AHNS}{\text{AHNS}}
\newcommand{\MWJO}{\text{MWJO}}

\newcommand{\Eq}[1]{Eq.~(\ref{#1})}
\newcommand{\Fig}[1]{Fig.~\ref{#1}}

\newcommand{\av}[1]{\left<{#1}\right>}

\begin{document}

\title{Scaling, Finite Size Effects, and Crossovers of the Resistivity
 and Current-Voltage Characteristics in Two-Dimensional Superconductors}

\author{Andreas Andersson}
\email{anan02@kth.se}

\author{Jack Lidmar}
\email{jlidmar@kth.se}

\affiliation{%
Theoretical Physics,
KTH Royal Institute of Technology,
AlbaNova,
SE-106 91 Stockholm,
Sweden}

\date{June 17, 2013}

\begin{abstract} 
  We revisit the scaling properties of the resistivity and the
  current-voltage characteristics at and below the
  Berezinskii-Kosterlitz-Thouless transition, both in zero and nonzero
  magnetic field.  The scaling properties are derived by integrating
  the renormalization group flow equations up to a scale where they
  can be reliably matched to simple analytic expressions.  The vortex
  fugacity turns out to be dangerously irrelevant for these quantities
  below $T_c$, thereby altering the scaling behavior.
  We derive the possible crossover effects as the current, magnetic
  field or system size is varied, and find a strong multiplicative
  logarithmic correction near $T_c$, all of which is necessary to account for
  when interpreting experiments and simulation data.
  Our analysis clarifies a longstanding discrepancy between the finite
  size dependence found in many simulations and the current-voltage
  characteristics of experiments.
  We further show that the logarithmic correction can be avoided by
  approaching the transition in a magnetic field, thereby simplifying
  the scaling analysis.
  We confirm our results by large scale numerical simulations, and
  calculate the dynamic critical exponent $z$, for relaxational
  Langevin dynamics and for resistively and capacitively shunted
  Josephson junction dynamics.
\end{abstract}


\maketitle

Fluctuation effects can be very strong in low-dimensional systems
and may radically alter the mean field picture of phase
transitions.  A well known example is that of two-dimensional (2D)
superfluids or superconductors, where phase fluctuations of the
complex order parameter $\psi = \psi_0 e^{i\theta}$ destroy long
range order at all nonzero temperatures.
Despite this, a superfluid/superconducting phase with algebraic
order, finite superfluid stiffness, and zero resistivity,
still exists at low temperature.  This is separated from the high
temperature disordered phase by a transition -- the
Berezinskii-Kosterlitz-Thouless (BKT) transition -- caused by the
thermal unbinding of vortex-antivortex
pairs~\cite{Berezinskii1,*Berezinskii2,KosterlitzThouless,MinnhagenRMP}.
The properties of 2D superconductors have been studied intensely in
recent years~\cite{Repaci1996,*Herbert1998,*MiuMiu2006,*Mondal2011,*Baturina,Strachan2003}
and continue to receive much interest due to the relevance for cuprate
superconductors with their layered structure.
Furthermore, advances in fabrication enable studies of single or few
atomic layer thick superconductors, which offer great potential for
precise tests against theories and simulations~\cite{Reyren,*Logvenov2009,*Ye2010}.
In this paper we explore the possible scaling behaviors and crossover
effects that may occur as a function of current, magnetic field, and
system size.  These results are confirmed by numerical simulations and
used for an accurate determination of the dynamic critical exponent
for two different equations of motion.

Transport measurements are perhaps the best way to experimentally
study the properties of 2D superconductors.  One of the hallmarks of
the BKT transition is the nonlinear current-voltage (IV) characteristics
$E \sim J^{a(T)}$ at and below $T_c$, with a temperature dependent
exponent $a(T)$~\cite{AHNS1,*AHNS2,HalperinNelson}.
The exact form of the temperature dependence of the exponent $a(T)$
has been subject to some debate~\cite{MWJO}.
According to the conventionally accepted theory
developed by Ambegaokar, Halperin, Nelson and Siggia (AHNS),
$a(T)= a_\AHNS = 1 + 2\pi \mathcal J(T)/2 T$, where
$\mathcal J(T) = \hbar^2 \rho_s(T)/2m$ is the superfluid stiffness and
$\rho_s(T)$ the (fully renormalized) superfluid areal
density~\cite{AHNS1,*AHNS2,HalperinNelson}.  This result has been
contested by Minnhagen \textit{et al.} (MWJO)~\cite{MWJO} who arrived at the
alternative expression $a(T) = a_\MWJO = 2\pi \mathcal J(T)/T - 1$
using scaling arguments.
Both yield $a=3$ at the transition $T_c = \pi\mathcal J(T)/2$.
Alternatively one may try to describe the data using a
Fisher-Fisher-Huse (FFH) scaling formula~\cite{FisherFisherHuse1991}
\begin{equation}							\label{eq:FFH}
  E = J \xi^{d-2-z} \mathcal E ( J \xi^{d-1}/T) ,
\end{equation}
where $\mathcal E(\cdot)$ is a scaling function and $\xi$ the
correlation length.  This leads also to a power-law, but leaves $a = z+1$
as a free fitting parameter related to the
dynamic critical exponent $z$ ($d=2$ is the dimension).
In 2D, however, fits of experimental data to \Eq{eq:FFH} easily give
surprisingly large values $a \gtrsim 6$~\cite{Pierson1999},
although more reasonable values $\approx 3$ have also been
obtained~\cite{Strachan2003}.
This, however, highlights the difficulty in using \Eq{eq:FFH} without
additional assumptions.  In any case it remains challenging to decide
which of the scenarios described above is correct based only on
experiments.
One may instead resort to computer simulations to try to settle the
controversy.
Usually, simulation data are analyzed using finite size scaling
formulas based on \Eq{eq:FFH}, with the diverging correlation length $\xi$
cut off by the system size $L$, yielding $E \sim J L^{1-a}$ for small $J$.
Most~\cite{MedvedyevaKimMinnhagen,KimMinnhagenOlsson,JensenKimMinnhagen,MinnhagenKimGronlund,Weber1996}
(but not all~\cite{SimkinKosterlitz,ChenTangTong,*TangChen})
simulation studies appear to favor the value $a_\MWJO$.
Interestingly, Refs.~\onlinecite{ChenTangTong,*TangChen}, obtain
agreement with both the AHNS and MWJO expressions in different regimes
and for different boundary conditions.
At the same time, the validity of the FFH scaling formula~\Eq{eq:FFH}
is still an open question, as is the scaling behavior in the presence
of an applied magnetic field.

The main contribution to the scaling behavior of the resistivity and
IV characteristics comes from the free vortex density $n_F$ of unbound
vortex pairs.  These can be either thermally excited or induced by an
applied magnetic field or a current.
Since only the motion of free vortices dissipate energy, the
resistivity should be proportional to the free vortex density
\begin{equation}								\label{eq:Drude}
  \rho = \Phi_0^2 \mu_v n_F ,
\end{equation}
where $\Phi_0$ is the flux quantum and $\mu_v \approx 2\pi\xi_0^2\rho_n
/ \Phi_0^2$ is the Bardeen-Stephen vortex
mobility.

Conventionally, 
the free vortex density $n_F = n_F^+ + n_F^-$ is calculated from a
rate equation~\cite{AHNS1,*AHNS2}
\begin{equation}									\label{eq:rate}
  \frac {dn_F^\pm}{dt} = \Gamma - \lambda n_F^+ n_F^-,
\end{equation}
where $\Gamma = \lambda \zeta^2 e^{-U_\text{eff}/T}$ is the pair
generation rate and $\lambda$ the recombination rate.  Here $\zeta =
e^{-E_c/T}$ is the vortex fugacity, and $E_c \sim \mathcal J$ the
vortex core energy.  The potential barrier to overcome in order to
create a pair of free vortices has two terms, one which depends
logarithmically on
their separation $r$, and one with a linear dependence due to the
applied current $U_\text{eff}(r) \approx 2\pi\mathcal J \ln(r/a_0) - J
\Phi_0 r$, where $a_0 \approx \xi_0$ is a short distance cutoff of the
order of the Ginzburg-Landau coherence length.
(From now on we set $a_0=1$.)
Optimizing gives $r^* \approx 2\pi \mathcal J/\Phi_0 J$
and $U_\text{eff} = U_\text{eff}(r^*) \approx - 2\pi\mathcal J \left[
  \ln (J \Phi_0 a_0/ 2\pi \mathcal J) + 1\right]$.
The stationary solution to \Eq{eq:rate} gives
\begin{equation}								\label{eq:nFz}
  n_F = 2 \zeta e^{-U_\text{eff}/2T} \sim 2 \zeta J^{2\pi \mathcal
  J / 2 T},
\end{equation}
and, with $E = \rho(J) J$, the result $a=a_\AHNS$.

There are several ways in which the above picture may need to be
modified.
First, interactions between vortices except those constituting the
pair are completely neglected.  Screening of the vortex interaction
from bound vortex-antivortex pairs can be taken into account by using
the fully renormalized value of the stiffness $\mathcal J(T)$ in place
of the bare one.
In a finite system the vortices may enter and exit the
system at the boundaries and \Eq{eq:rate} will acquire more terms
describing these processes.
Accounting for a realistic geometry and nonuniform current
distribution can lead to a rather complicated
behavior~\cite{GurevichVinokur}.
In simulations one usually avoids surface effects by using periodic
boundary conditions (PBC).
Finite size effects, however, become visible when $r^* = 2\pi \mathcal
J / \Phi_0 J \gtrsim L$, leading to a crossover to ohmic behavior at
low currents, with a characteristic size dependent resistivity.
Another issue is that the rate equation \eqref{eq:rate} presumes that
density fluctuations are small, which is true for large systems, but
not for small enough systems with area $L^2 \lesssim 1/n_F$.  In the
latter regime the constraint of vortex-antivortex neutrality (enforced
when using PBC~\footnote{By PBC we mean, here and in the following,
  any boundary condition which enforces vortex neutrality.}) instead
leads to $\Gamma/\lambda = \av{n_F^+ n_F^-} \approx \av{n_F^\pm}^2 +
L^{-2} \av{n_F^\pm}$, which is dominated by the second term, i.e.,
\begin{equation}								\label{eq:nFz2}
  n_F \sim 2 L^2 \zeta^2 e^{-U_\text{eff}/T} ,
\quad (\text{PBC and } L^2 n_F \lesssim 1) .
\end{equation}
The same expression follows from a low fugacity expansion of the
neutral Coulomb gas, which only involves even powers of $\zeta$.
Also note that an applied perpendicular magnetic field $B$ will lead to a net density of free
vortices $\Delta n = n_F^+ - n_F^- = B/\Phi_0$, such that
\begin{equation}								\label{eq:nFB}
n_{F}^2 = {\Delta n^2 + 4 n_F^+ n_F^-} \approx {\Delta
  n^2 + 4 \zeta^2 e^{-U_\text{eff}/T}} .
\end{equation}

A more systematic approach to take into account interaction effects,
is to first integrate the renormalization group (RG) flow up to the
scale where one of the coupling constants becomes large of $O(1)$ and
only then match the theory to simple approximate expressions similar
to the ones discussed above.
The RG flow equations are most easily expressed in the Coulomb gas
language using the rescaled temperature and fugacity variables,
$ x = 1 - \frac {\pi \mathcal J}{2 T}$, $y = 2\pi \zeta$.
To lowest order in $x$ and $y$ they read~\cite{Kosterlitz,MinnhagenRMP}
\begin{align}
\frac{dx}{d \ell} &= 2y^2 , &
\frac{dy}{d \ell} &= 2xy ,  \label{eq:RG}
\end{align}
where $\ell = \ln b$ is the logarithm of the scale factor $b$.
%
%
The resulting RG flow obeys $x^2 - y^2 = C^2$, where
\begin{equation}								\label{eq:C}
  |C| =\sqrt{|x_0^2-y_0^2|} \approx c \sqrt{|T_c-T|}
\end{equation}
is a constant determined by the initial conditions.
Below $T_c$ we have $C^2 > 0$ and the RG flow ends up on a critical
line $x = - C < 0$, $y = 0$ as $\ell \to \infty$.
Above $T_c$, $C^2 < 0$ and the flow will eventually diverge to $+\infty$.
The BKT transition occurs at $T=T_c$, where the flow follows the
separatrix $x = -y$.
In order to describe the various crossovers we need the explicit
solutions~\cite{Kosterlitz},
$y(\ell) =  C/{\sinh(2C(\ell-\ell_0))}$ for $T<T_c$,
$y(\ell) = 1 / {(2\ell - 2\ell_0)}$ for $T = T_c$, and
$ y(\ell) = -{ |C| } / {\sin(2|C|(\ell - \ell_0))}$ for $T > T_c$.
In terms of 
$b = e^\ell$ we have
\begin{align}
  \label{eq:RGsolution}
  y(b) &= \frac {2 C (b/b_0)^{-2C}} {1 - (b/b_0)^{-4C}}, & (T < T_c), \\
  y(b) &= \frac 1 {2\ln(b/b_0)} , & (T = T_c),
\end{align}
where $b_0 = e^{\ell_0}$ is fixed by the initial conditions.
Near $T_c$, where $|C| \lesssim y_0$,
we have to a good approximation $c^2 \approx 4 y_0 /\pi \mathcal J$,
$T_c \approx \pi \mathcal J / 2(1+y_0)$, $\ell_0 \approx - 1/2y_0$.
Further below $T_c$, where $C \gtrsim y_0$, we have instead 
$C \approx -x_0$,
so that
\begin{align}								\label{eq:RGsolutionfarbelow}
  y(b) &\approx y_0 b^{-2C} , & (T \ll T_c) .
\end{align}
Note also that $C = -x(b \to \infty) = \pi \mathcal J_R(T)/2T - 1$ is directly
related to the fully renormalized superfluid stiffness $\mathcal J_R(T)$.

\begin{figure*}
  \begin{center}
    \includegraphics[width = 0.75\linewidth]{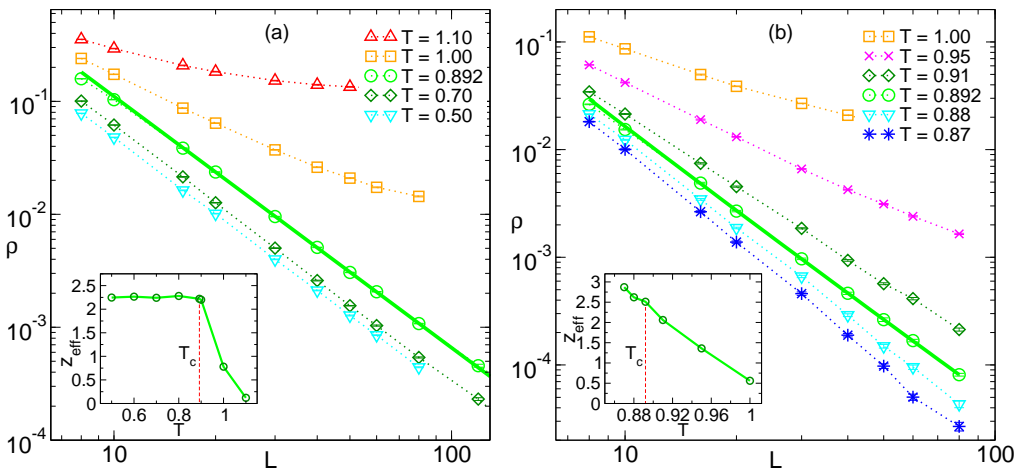}%
    \caption{(Color online) \emph{Langevin dynamics.}
      (a) Resistivity $\rho$ vs system size $L$ at different temperatures
      in a magnetic field $B = \Phi_0/L^2$.
      The dotted lines are guides for the eyes, and the full green curve at $T = T_c$ is a $\chi^2$-fit (using
      $L=16$--$120$) to the power-law $\rho \sim L^{-z}$, giving
      $z = 2.22$.
      (b) As in (a), but for zero magnetic field.
     The full green curve at $T = T_c$ is a $\chi^2$-fit (using
     $L=16$--$80$) to $\rho \sim
     L^{-2.22} / (\ln L -\ell_0)^2$ 
     with fixed $z = 2.22$, giving $\ell_0 = -2.71$.
     Insets: The effective exponent $z_\text{eff}$ vs temperature, obtained from power-law
     fits. Note how $z_\text{eff}$ is almost constant below $T_c$ in (a).
      \label{fig:langevin@f_onevortex_rho_vs_L}
    }
  \end{center}
\end{figure*}

The free vortex density, being the vortex density which remains after
the elimination of all bound pairs, is only rescaled by the RG
transformation and therefore has scaling dimension 2, i.e,
$n_F \sim b^{-2}$.
As a function of system size $L$, magnetic flux density $B$, current $J$,
$x$, $y$, and 
possibly other perturbations
it therefore transforms as
\begin{multline}								\label{eq:nF-scaling}
  n_F(x_0,y_0, L,B,J,\ldots) = \\ b^{-2} n_F(x(b),y(b), Lb^{-1}, Bb^2,Jb,\ldots)
\end{multline}
under the RG.
A similar equation holds for the resistivity \Eq{eq:Drude}.
Most theories assume that the vortices undergo ordinary diffusion
However, we are not aware of any argument which
prevents the renormalization of the vortex mobility $\mu_v$
in \Eq{eq:Drude}.  Hence, we allow for an anomalous
dimension $\mu_v \sim b^{2-z}$, with a dynamic critical exponent $z$
not necessarily fixed to 2, such that the resistivity transforms as
\begin{multline}								\label{eq:rho-scaling}
  \rho(x_0,y_0, L,B,J,\ldots) = \\
  b^{-z} \rho(x(b),y(b), Lb^{-1}, Bb^2,Jb,\ldots).
\end{multline}
An FFH scaling formula follows from \Eq{eq:rho-scaling} \emph{if}
$\rho$ flows smoothly to a nonzero constant as $b \to \infty$.
This is the case above $T_c$, where the flow must be stopped at a
scale when $x \sim y \sim O(1)$, 
yielding the Debye-Hückel expression $n_F \approx 1/2\pi \xi_+^2$,
where $\xi_+ \sim\exp({\pi/2c\sqrt{T-T_c}})$ is the correlation length
above $T_c$.
This is, however, \emph{not} the case in zero magnetic field
at and below $T_c$, where $y = 2 \pi \zeta \to 0$, because $n_F$ vanishes 
in this limit.
In other words, the fugacity is dangerously irrelevant for $n_F$ and
$\rho$ in this case.
Instead the right hand side of
Eqs.~\eqref{eq:nF-scaling}-\eqref{eq:rho-scaling}
must be matched to one of Eqs.~\eqref{eq:nFz}-\eqref{eq:nFB}.
At the matching scale $b$ the barrier $U_\text{eff}$ in \Eq{eq:nFz} or
\eqref{eq:nFz2} has reduced to zero, and we are left with three different
possibilities:
In zero magnetic field $n_F(b) \sim y(b)$ or $y^2(b)$ depending on boundary
conditions and system size, while for nonzero field
$n_F(b) \approx \sqrt{B^2b^4/\Phi_0^2 + y^2(b)/\pi^2}$.
This will turn out to have profound consequences for the scaling of
many quantities.

We first discuss the finite size scaling of the linear resistivity in
zero magnetic field.  The RG flow must then be stopped at $b = L$.
Under the RG all length scales, including the system size, shrink by a
factor $b$ so that the effective system size becomes $L' = L/b = 1$.
The system must therefore be matched to \Eq{eq:nFz2} when using
periodic boundary conditions, or to \eqref{eq:nFz} when using open
boundary conditions.
For PBC we thus get $\rho(L) \sim L^{-z} y^2(L)$, and by using
Eqs.~\eqref{eq:RGsolution}-\eqref{eq:RGsolutionfarbelow},
the limiting cases
\begin{align}									\label{eq:FSS}
  \rho(L) \sim
  \begin{cases}
    L^{-z + 4 - 2\pi \mathcal J_R/T}, & (L \gtrsim \xi_-), \\
    L^{-z} / \ln^2(L/b_0), & (L \lesssim \xi_-) ,
  \end{cases}
\end{align}
where $\xi_- \approx \exp({1/2C}) \approx \exp({1/2c\sqrt{T_c-T}})$ is
the correlation length below $T_c$, defined as the scale on which
$x(b)$ has approximately reached its asymptotic value $-C$.
The power-law appearing in this expression agrees with the finite size
scaling of MWJO~\cite{MedvedyevaKimMinnhagen} if one assumes $z=2$.
On the other hand, for open boundary conditions $\rho(L) \sim
L^{-z}y(L)$, or
\begin{align}									\label{eq:FSS-open}
  \rho(L) \sim
  \begin{cases}
    L^{-z + 2 - \pi \mathcal J_R/T}, & (L \gtrsim \xi_-), \\
    L^{-z} / \ln(L/b_0), & (L \lesssim \xi_-),
  \end{cases}
\end{align}
which, for $z=2$, would be consistent with the AHNS scaling.
The finite size scaling at $T_c$, where $\xi_- = \infty$, has in both
cases, strong \emph{multiplicative} logarithmic corrections.

\begin{figure*}[t]
  \begin{center}
    \includegraphics[width = 0.75\linewidth]{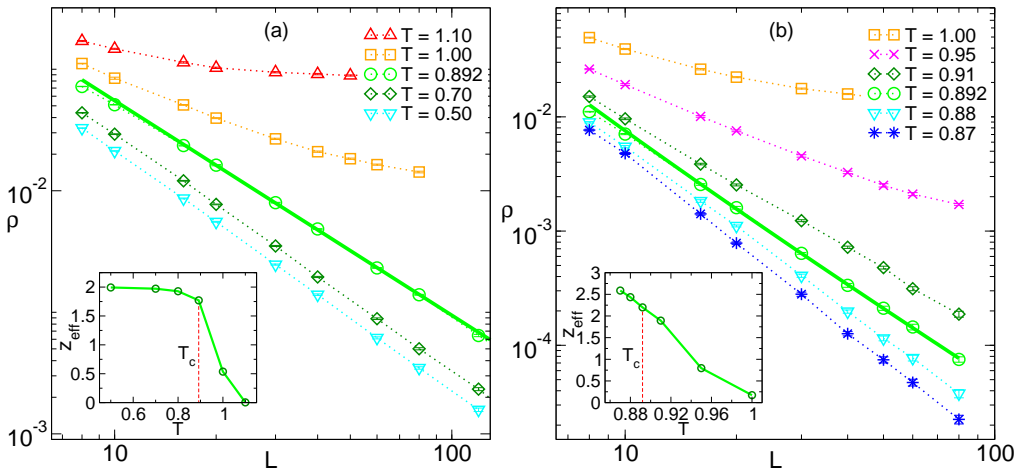}
    \caption{(Color online)
      \emph{Overdamped RCSJ dynamics} {(Stewart-McCumber $\beta_C =2\pi I_c R^2C/\Phi_0 = 0.25$).}
      (a) Resistivity $\rho$ vs system size $L$ at different
      temperatures
      in a magnetic field $B = \Phi_0/L^2$.
      The full green curve at $T = T_c$ is a $\chi^2$-fit (using
      $L=16$--$120$) to the power-law $\rho \sim L^{-z}$, 
      giving 
      $z = 1.77$.
      (b) As in (a), but for zero magnetic field.
      The full green curve at $T = T_c$ is a $\chi^2$-fit (using
      $L=16$--$80$) to
      $\rho \sim L^{-1.77} / (\ln L - \ell_0)^2$ with fixed $z=1.77$,
      giving 
      $\ell_0 = -1.33$.
      Insets: As in \Fig{fig:langevin@f_onevortex_rho_vs_L}.
      \label{fig:rcsj@f_onevortex_rho_vs_L}
    }
  \end{center}
\end{figure*}

The situation in a nonzero magnetic field is different.
The magnetic field is a relevant perturbation, which destroys
superconductivity by introducing a finite density of free vortices
even at low temperature.
We can, however, still approach the transition by scaling down the
magnetic field with the system size, holding $BL^2 = N \Phi_0$, the
net number of flux quanta, fixed.
(This is easy in a simulation, but more difficult in an experiment.)
Consider, e.g., the case $N=1$.
Stopping the RG flow at $b \sim L = \sqrt{\Phi_0/B}$
and matching to \Eq{eq:nFB} then gives
$\rho \sim L^{-z} \sqrt{1 + y^2(L)/\pi^2}$.
The leading scaling behavior thus remains a temperature independent
power-law with exponent $z$ in contrast to the zero field case (with a
weak \emph{additive} correction decreasing with system size).

The finite size scaling formulas derived above are well-suited for the
analysis of numerical simulations.  We have performed simulations of
the 2D XY model, defined by the Hamiltonian
$H = -\mathcal J \sum_{\av{ij}} \cos(\theta_i - \theta_j-A_{ij})$,
using two types of dynamics, relaxational Langevin dynamics and
resistively and capacitively shunted Josephson junction (RCSJ)
dynamics (in the overdamped limit)~\cite*{[Details of the simulation
  methods can be found in ]Andersson2011}.
The resistivity was calculated from the equilibrium voltage
fluctuations using a Kubo formula, with a sampling time
of $10^6$--$10^8$ time units per datapoint.
For an accurate determination of $z$ we
apply a weak magnetic field $B = \Phi_0/L^2$ so that the system
contains exactly one vortex irrespective of system size.
This minimizes the influence of the logarithmic correction near $T_c$,
allowing us to fit the data for $T \le T_c$ to the simple scaling law
$\rho(L) \sim L^{-z}$.
We plot, in \Fig{fig:langevin@f_onevortex_rho_vs_L}(a), $\rho$ vs $L$
calculated using Langevin dynamics on
a log-log scale for a range of temperatures
including $T_c$ ($T_c \approx 0.892 \mathcal J$~\cite{Olsson}).
The data at and below $T_c$ do indeed follow a power-law with a
temperature independent exponent $z \approx 2.22 \pm 0.05$.
In contrast, the zero field data shown in
\Fig{fig:langevin@f_onevortex_rho_vs_L}(b) follow different
power-laws at different temperatures.
Right at $T_c$ the data is very well fitted by \Eq{eq:FSS} with $z$
fixed to $2.22$.  
The value of $\ell_0 = \ln b_0 \approx -2.7$ obtained by the
fit compares well with the theoretical estimate $\ell_0 \approx
-1/2y_0 \approx -2$ obtained using the XY value
$y_0=2\pi e^{-E_c/T}$, with $E_c \approx \pi^2\mathcal J/2$.
Without knowing about the logarithmic correction one would fit the
data at $T_c$ to a pure power-law and draw the wrong conclusion.  For
our data this would give an effective exponent $z_\text{eff} \approx
2.54$, appreciably
different from the true $z$.

Figure \ref{fig:rcsj@f_onevortex_rho_vs_L}
shows similar plots for RCSJ dynamics.  The resistivity for a system
with exactly one vortex again follows a power-law, but this time with
$z=1.77 \pm 0.05$ at $T_c$.  In zero field the data
is well fitted to \eqref{eq:FSS} using the same $z$, with $\ell_0
\approx -1.33$ again in rough agreement with expectations, whereas a pure
power-law fit would give a too large exponent $z_\text{eff} \approx
2.2$.

The values $z \approx 2.22$ and $z \approx 1.77$ for Langevin and RCSJ
dynamics, respectively, are close to, but significantly different from the
conventional value 2, and correspond either to subdiffusive ($z>2$) or
superdiffusive ($z<2$) vortex motion.

The scaling behavior below $T_c$ differs considerably in zero and
nonzero magnetic field.  As seen in the insets of
Figs.~\ref{fig:langevin@f_onevortex_rho_vs_L} and
\ref{fig:rcsj@f_onevortex_rho_vs_L} the resistivity with
$B=\Phi_0/L^2$ follows a power law with practically
temperature-independent exponents in stark contrast to the zero field
case.
Previous finite size scaling studies of $\rho(L)$ (or $E(J,L)$ in
the ohmic regime) in zero field
have obtained a temperature-dependent power-law exponent below $T_c$
in good agreement with the MWJO
prediction~\cite{KimMinnhagenOlsson,JensenKimMinnhagen,MinnhagenKimGronlund,Weber1996,ChenTangTong,*TangChen},
which is not surprising given \eqref{eq:FSS} and the smallness of
$z-2$.

In a large or infinite system at zero magnetic field,
the RG flow must be stopped at a scale dictated by the applied
current, i.e., when $Jb \approx J_0 =
2\pi \mathcal J /\Phi_0$.
At this scale the matching condition is $n_F \sim y$ and
the nonlinear resistivity $\rho(J)
\sim J^z y(b \approx J_0/J)$ obtains from
Eqs.~\eqref{eq:RGsolution}-\eqref{eq:RGsolutionfarbelow}.  We have the
limiting cases
\begin{equation}								\label{eq:IV}
  \rho(J) = \frac E J \sim
  \begin{cases}
    J^{z + \pi \mathcal J_R(T)/T - 2}, \quad & J_0/J \gtrsim \xi_- ,\\
    J^{z}/\ln (J_0/Jb_0), \quad & J_0/J \lesssim \xi_- .
  \end{cases}
\end{equation}
The power-law behavior at low currents below $T_c$ is in agreement with the AHNS
value if one assumes $z=2$.  Close to $T_c$ we find a strong
multiplicative logarithmic correction.
The crossover to the finite size induced ohmic behavior in \Eq{eq:FSS} or
\eqref{eq:FSS-open} happens when $J L \lesssim J_0$.  In addition one
expects a high-current crossover to an ohmic regime when $J \gtrsim
J_0$.

In the PBC case it is also possible to have an intermediate regime
where the matching is still done at a scale $b \approx J_0/J$, but the
effective system size is small enough that $n_F(b) (L/b)^2 \lesssim
1$, so that $n_F \sim y^2$.  This would give
\begin{align}								\label{eq:intermediate}
  \rho(J,L) & \sim L^2 J^{z -2 + 2\pi \mathcal J_R /T}, & \frac{J_0}{J} \lesssim L \lesssim
  \left(\frac{J_0}{J}\right)^{\pi\mathcal J_R / 2T} 
\!\!\!\!\!\!\!\!\! .
\end{align}
Such an intermediate scaling regime was previously proposed in
Ref.~\onlinecite{ChenTangTong,*TangChen}, using an entirely different
approach.

To summarize, we have obtained a coherent picture of the scaling
behavior and crossover effects of the (nonlinear) resistivity near and
below the BKT transition,
Eqs.~\eqref{eq:FSS}--\eqref{eq:intermediate}. 
The finite size results depend sensitively on the boundary conditions
and on whether a magnetic field is present or not.
In the limit of large systems the IV exponent agrees with the AHNS
result, with the modification that we allow for the possibility that
$z \neq 2$.
For PBC, on the other hand, the finite size scaling agrees with MWJO.
Our simulations suggest that $z$ differs from $2$ and moreover that Langevin and
RCSJ dynamics belong to different dynamic universality
classes~\cite{HohenbergHalperin}.
From a practical point we found it important to take into account the
logarithmic correction near $T_c$ when analyzing finite size data.
The same should hold true for experimental finite current data.
Note, however, that to make quantitative comparisons with experiments
it may be important to consider effects of inhomogeneity and pinning,
and to make realistic estimates of the temperature dependence of the
bare parameters $\mathcal J$, $y$, e.g., using Ginzburg-Landau
theory~\cite{BenfattoCastellaniGiamarchi}.
Finally, it should be noted that the only assumptions needed in our
analysis is the low fugacity behavior of the zero magnetic field
resistivity $\rho \sim y$ or $y^2$.  It is highly likely that other
quantities may be affected in similar ways.


We thank M.~Wallin for comments on the manuscript.
This work was supported by the Swedish Research Council (VR) through
grant no.\ 621-2007-5138 and the Swedish National Infrastructure for
Computing (SNIC 001-10-155) via PDC.

\bibliography{bkt}

\end{document}